\documentclass[journal]{IEEEtran}
\IEEEoverridecommandlockouts
\usepackage{cite}
\usepackage{amsmath,amssymb,amsfonts}
\usepackage{algorithmic}
\usepackage{graphicx}
\usepackage{textcomp}
\usepackage{xcolor}
\usepackage{siunitx}
\usepackage[T1]{fontenc}
\usepackage{subfigure}
\usepackage{comment}
\usepackage{lipsum}

\usepackage{color,soul}

\def\BibTeX{{\rm B\kern-.05em{\sc i\kern-.025em b}\kern-.08em
    T\kern-.1667em\lower.7ex\hbox{E}\kern-.125emX}}

\begin{document}

\title{Power Line Communication vs.~Talkative Power Conversion: A Benchmarking Study}

\author{\mbox{ }
\thanks{\mbox{ }}%
\thanks{\mbox{ }}}

\author{Peter A.~Hoeher, Yang Leng, Maximilian Mewis, and Rongwu Zhu
}

\renewcommand{\baselinestretch}{1} 

\maketitle

\begin{abstract}
The convergence of energy transmission and data communication has become a key feature of decentralized energy systems across a broad spectrum of voltage/power ranges, including smart grid applications and cyber-physical power systems. This paper compares two distinct approaches: Power Line Communications (PLC) and Talkative Power Conversion (TPC). While PLC leverages existing power infrastructure for data transmission by using external data transmitters and receivers, TPC integrates communication capabilities directly into power electronic converters. We present their technical foundations and applications, benchmark their strengths and bottlenecks, and outline future research directions regarding TPC that could bridge the gap between power and communication technologies.
\end{abstract}



\section{Introduction}\label{Introduction}
Modern energy systems demand seamless integration of power delivery and data transfer to support network intelligence, device interoperability, data and energy flow, and energy efficiency -- both in wireline and wireless applications. Conventional approaches often treat power and communication as separate domains, but simultaneous information and power transfer techniques like Power Line Communication (PLC) and Talkative Power Conversion (TPC) bridge this gap \cite{Liserre2023}. PLC exploits power lines as data channels, whereas TPC uses power converters as dual-purpose devices for energy conversion and information modulation \cite{Popovski2017, He2020}. This paper targets professionals in the fields of power electronics (PE) and communication engineering (COM) who wish to gain a deeper understanding of the fundamentals and a comparison of PLC and TPC.

PLC has been designed ``for the grid and through the grid'' \cite{Galli2011}. In PLC, simultaneous information and power transfer is achieved by superimposing a modulated carrier signal onto an existing power line \cite{Lampe2016, Ndjiongue2019}. In most PLC standards, rate-adaptive high-order multicarrier modulation schemes are used. Besides power lines, some PLC standards are applicable to twisted pairs and coax cables as well. Digital signal processing (DSP) and amplification are implemented in external modems that are connected to the power line by inductive or capacitive coupling \cite{Berger2015}. Narrowband PLC systems are widely applied for grid telemetry, load control, and automated meter reading. On the other side, broadband PLC systems are popular for home/industrial automation, broadband Internet and infotainment, as they offer data rates up to the Gbps range. 

TPC introduces a paradigm shift where power converters not only regulate energy flow but also embed data signals within power waveforms \cite{Liserre2023}. This is achieved through joint power/data modulation techniques that enable converters to ``talk'' by superimposing a communication signal onto the direct current (DC) or alternating current (AC) waveform delivered by a switched-mode power electronics converter (PEC). Originally, TPC has been proposed for grid applications \cite{Stefanutti2008}. Later, ``power talk'' \cite{Popovski2017} and ``talkative power'' \cite{He2020} have been extended to battery management systems, electrical machine systems, wireless charging, and optical applications (e.\,g., visible light communication) \cite{Liserre2023}. TPC technology is low-cost, reliable, and sustainable, as no additional PE components are to be used. In recent years, there has been a consistent increase in the number of publications related to TPC aspects \cite{Hoeher2021, Han2022, Wang2023, Chen2023, Mousavi2024, Leng2025}.

Although PLC is widely used worldwide and TPC is approaching maturity after about a decade of research, a solid comparison is apparently lacking. Moreover, TPC is not well known in the communications community, although its roots are in PE and COM engineering \cite{Stefanutti2008, Popovski2017}. This paper intends to fill the gap. Our main contributions are summarized as follows:
\begin{itemize}
\item We provide a comprehensive comparison between PLC and TPC, evaluating their respective advantages, limitations, and potential applications in modern electrical networks. 
\item We suggest a roadmap for future research directions at all layers. Among the proposals is a hybrid PLC-TPC system, which is being introduced here for the first time. 
\end{itemize}

Besides PLC and TPC, many other simultaneous information and power transfer schemes have been invented, particularly for low-power applications. These include bus systems (e.\,g., CAN, PROFIBUS, HyperTransport, InfiniBand, Thunderbolt), Power over Ethernet (PoE), wireless charging, and wireless powered networks \cite{Ng2019}. These options are beyond the scope of our benchmarking study on PLC and TPC, however. 

\section{Introduction to Power Line Communication}\label{PLC}

\begin{table*}[t]
\caption{Key parameters of selected narrowband PLC standards (top) and broadband PLC standards (bottom).}
\centering
\begin{tabular}{l|llll}
\hline
\textbf{Name }     & \textbf{Spectrum}            & \textbf{Max.~data rate (PHY)}  & \textbf{Technology}  & \textbf{Application notes} \\
\hline
PRIME              & 42–89~kHz (CENELEC A)        & 128.6~kbps           & OFDM (FFT-based)          & Smart metering (Europe)    \\
G3-PLC             & 10–490~kHz (FCC)             & 300~kbps             & OFDM (ROBO mode)          & Global smart grid          \\
IEEE 1901.2        & 10–490~kHz (FCC/CENELEC)     & 500~kbps             & OFDM/FSK                  & Smart grid/utility networks\\
IEC 61334          & typ.~10-100~kHz (CENELEC A)  & 2.4~kbps - 4.8~kbps  & Spread-FSK                & Utility automation     \\
\hline
HomePlug AV        & 2-30~MHz                     & 200~Mbps             & OFDM (FFT-based)          & Home networking           \\
HomePlug AV2       & 2-86~MHz                     & 1.5~Gbps (MIMO)      & OFDM (FFT-based)          & High-speed networking     \\
HomePlug GP        & 2-30~MHz                     & 10~Mbps              & OFDM (FFT-based)          & Smart grid/IoT            \\
ITU-T G.hn         & 2-200~MHz (PL), 350-2850~MHz (coax) & 1-2~Gbps      & OFDM (FFT-based)          & Multi-media (power line/coax) \\
IEEE 1901 FFT      & 2-50~MHz                     & 500~Mbps             & OFDM (FFT-based)          & Interoperable broadband   \\
IEEE 1901 HD-PLC   & 2-50~MHz                     & 420~Mbps             & OFDM (Wavelet-based)      & Industrial/robust communication  \\
\hline 
\end{tabular}
\label{nbPLC_vs_bbPLC}
\end{table*}

\subsection{Operating Principle of PLC}
\begin{figure}[h]
\centering
\includegraphics[width=0.485\textwidth]{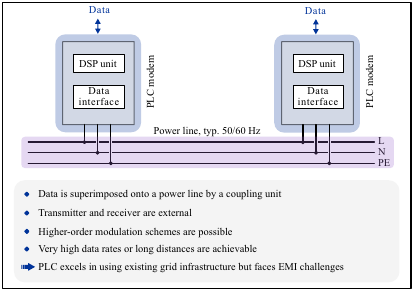}
\caption{Structure and main takeaways about PLC.}
\label{PLCsummary}
\end{figure}

PLC technology utilizes the existing electrical power grid as a communication medium by injecting high-frequency modulated signals onto the grid \cite{Lampe2016}. This approach enables bidirectional data transmission without requiring dedicated wiring infrastructure, see Fig.~\ref{PLCsummary}, making it a cost-effective solution for smart grid applications, home automation, and broadband access. A critical component of PLC systems is the coupling mechanism, which facilitates the interface between external communication devices and the power line. Inductive or capacitive coupling methods are commonly employed for this purpose. Inductive coupling uses magnetic fields to transfer signals via transformers or coils. Capacitive coupling, on the other hand, leverages electric fields through coupling capacitors. Both methods require careful circuit design to ensure impedance matching, minimize signal attenuation, and suppress interference from grid-connected equipment.

The high-frequency modulated transmit signal superimposes the grid voltage. 
Most PLC systems have been deployed over the standard 50/60 Hz AC grid. 
However, there is a growing interest in PLC over DC networks, particularly for microgrids \cite{Ndjiongue2019}. 
Applications such as in-vehicle grids are related to bus systems like the CAN bus, among others.
Another relevant application involves the communication between plug-in electric vehicles and their respective charging infrastructures. 
Additionally, direct current is transmitted between solar panels and battery packs, for instance. 

To overcome the challenges posed by the frequency-selective nature of power line channels, modern PLC systems frequently adopt orthogonal frequency-division multiplexing (OFDM) \cite{Lampe2016, Ndjiongue2019}. OFDM divides the signal bandwidth into multiple orthogonal subcarriers, enabling robust data transmission in environments plagued by impulsive noise, frequency-selective attenuation, and narrowband interference. Adaptive bit-loading techniques further optimize spectral efficiency by dynamically assigning quaternary or higher-order modulation schemes to individual subcarriers. 

The utilization of multiple-input multiple-output (MIMO) processing has further enhanced PLC performance by exploiting the inherent multi-wire structure of power lines (e.\,g., live (L), neutral (N), and protective earth (PE)). Delta-style, T-style and star-style inductive couplers  offer multiple input and output ports \cite{Berger2015}. MIMO-PLC systems are able to employ spatial multiplexing to transmit independent data streams across separate wires, significantly improving channel capacity and reliability. Additionally, spatial diversity techniques mitigate the impact of frequency-selective fading and impulsive noise by combining signals from multiple receiving paths. However, the strong correlation between power line channels and cross-talk between wires pose challenges for MIMO implementations, necessitating sophisticated DSP algorithms for channel estimation and interference cancellation.
In Fig.~\ref{PLCsummary}, main takeaways about PLC are summarized. 

\subsection{Classification of PLC and PLC Standards}
PLC can be broadly classified into two categories: Narrowband (NB) PLC and Broadband (BB) PLC \cite{Lampe2016, Ndjiongue2019}. Typically operating below 500~kHz, NB-PLC is optimized for low-data-rate applications such as remote metering, utility automation, and grid monitoring. Its robustness in long-range communications makes it a favorable choice for utility networks despite its limited bandwidth. BB-PLC exploits higher signal frequencies (up to 250~MHz) and higher-order modulation schemes to achieve higher data rates, enabling more bandwidth-intensive applications like Internet access, multimedia streaming, and advanced automation infrastructure. Compared to NB-PLC, this approach is more sensitive to noise and requires sophisticated interference cancellation and/or error-correction algorithms to combat interference and signal degradation. BB-PLC is only suitable for short ranges. 

TABLE~\ref{nbPLC_vs_bbPLC} depicts key parameters of some prominent NB- and BB-PLC standards. 
Since NB-PLC standards operate in the spectrum below 500~kHz, their data rates are typically limited to several hundred kbps. The most widely used standards in these applications are IEEE 1901.2, G3-PLC, and PRIME. Typical applications include low-speed diagnostic and control operations in smart grid applications (like automatic meter reading, grid monitoring, and fault detection) as well as industrial use cases. IEEE 1901.2 supports up to 72~kV grids. Additionally, IEC 61334 focuses on MV distribution networks (1-36~kV) for utility automation, load control, fault location, and grid management. 

Among the popular BB-PLC standards are HomePlug, ITU-T G.hn, and IEEE 1901.  
HomePlug is the trade name of the PLC family of standards developed by the HomePlug Powerline Alliance. Main applications of HomePlug~AV/AV2 are broadband Internet, high-definition video, digital music, and smart energy applications. HomePlug Green PHY (HomePlug GP) is a special simplified version designed for smart grid applications like smart electric meters, heating, ventilation, and air conditioning (HVAC) control as well as appliance monitoring, but also plug-in electric vehicles. GreenPHY uses up to 75~\% less energy than AV thanks to simplifications. 
G.hn is a standard of the ITU-T organization. The main applications are home networking, Internet access, and connecting video equipment. Besides PLC, G.hn also works with phone-line and cable TV wiring.
IEEE 1901 is related to HomePlug and ITU-T G.hn. The IEEE 1901 standard defines two broadband PLC techniques, whose physical layer (PHY) are based on FFT-based OFDM and on wavelet-based OFDM, respectively. The FFT-based PHY of IEEE 1901 is derived from HomePlug~AV technology, hence all HomePlug standards such as AV, AV2, and GP are interoperable with IEEE 1901 products. The wavelet PHY is promoted by the HD-PLC Alliance and targets industrial and robust communication. 

\subsection{Challenges and Progress of PLC}
Recent research has focused on overcoming challenges inherent to PLC, including impulse noise from electrical appliances and regulatory constraints that dictate transmission power and frequency bands. Innovations in adaptive modulation, channel coding, and digital signal processing have contributed to improved data throughput and reliability in various deployments. This includes MIMO processing \cite{Berger2015}, wave traps, multi-hopping and repeaters to improve the signal quality and/or the communication range. 

\section{Introduction to Talkative Power Conversion}\label{TPC}
\subsection{Operating Principle of TPC}
\begin{figure}[th]
\centering
\includegraphics[width=0.485\textwidth]{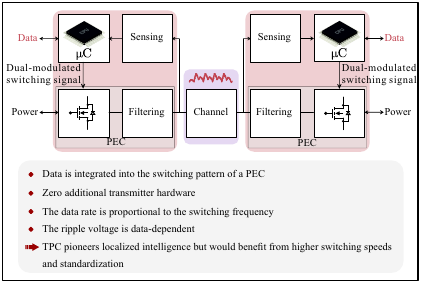}
\caption{Structure and main takeaways about TPC.}
\label{TPCsummary}
\end{figure}

In TPC, the PWM-type switching pattern the controls switched-mode PECs is modified to enable joint energy and data modulation, referred to as power/signal dual modulation \cite{Wang2023}. PWM waveforms are traditionally designed solely to regulate power flow by controlling the duty cycle of square-wave pulses to achieve desired output voltages or currents. In contrast, power/signal dual modulation introduces controlled perturbations of the PWM signal to embed information while maintaining the primary function of efficient energy conversion \cite{Liserre2023, Wang2023}. This approach leverages the inherent switching dynamics of PECs, particularly DC--DC converters \cite{Chen2023} and DC--AC inverters \cite{Leng2025}, to simultaneously transmit information without requiring additional communication hardware at the transmitter side \cite{Han2022}. The information is contained in the ripple voltage. Dual modulation can be modified by varying pulse widths, phase shifts, or switching frequencies within permissible limits that do not compromise the converter’s energy efficiency or output quality \cite{Hoeher2021, Wang2023}. However, this dual functionality introduces challenges, including trade-offs between modulation depth, switching frequency, data rate, and power quality, as well as electromagnetic radiation and signal distortion under dynamic load conditions. The data rate is proportional to the switching frequency and hence naturally limited. Recent advances like advanced converter topologies, adaptive modulation and channel coding schemes, multiple access schemes, advanced receiver algorithms, and machine learning address these issues \cite{Hoeher2021, Mousavi2024}. By integrating communication capabilities directly into power conversion hardware, dual modulation reduces system complexity, cost, and latency, paving the way for more interconnected and intelligent energy systems. This makes it predestined for cyber-physical power systems, i.\,e., for advanced frameworks that integrate physical power infrastructure (e.\,g., generators, transmission lines, loads) with cyber components (e.\,g., sensors, communication networks, control algorithms) to enable real-time monitoring, intelligent control, and resilient operation of electric grids or powered sensor networks \cite{Ng2019}. 
In Fig.~\ref{TPCsummary}, the structure of TPC is depicted and main takeaways about TPC are summarized. 

\subsection{Challenges and Progress of TPC} 
{\em Switching frequency constraint:} The data rate in PE applications is typically limited by the converter switching frequency, which usually ranges between 20~kHz and 100~kHz. This limitation can be relaxed by adopting wide-bandgap semiconductor materials such as gallium nitride (GaN) and silicon carbide (SiC).

{\em Two-level switching constraint:} Since any switching signal is inherently binary (ON or OFF), this restriction narrows the choice of viable modulation schemes and caps the achievable data rate. Nevertheless, advanced power converters incorporating multiple switching units can enhance the degrees of freedom for data modulation, enabling MIMO signal processing.

{\em Trade-off between ripple voltage and power quality:} From a PE point of view, the ripple voltage is undesired and should be as small as possible. Vice-versa, from a COM point of view the ripple voltage carries the information and therefore should be large enough for reliable data detection at the receiver side. Possible counter measures include line coding, error correction coding, adaptive modulation taking the environment into account, and relaying. 

{\em Interference challenges:} Signal quality degradation due to interference from interconnected converters and inverters necessitates the use of multiuser access strategies, interference cancellation techniques, and error-correction codes to maintain reliable communication.

{\em Standardization deficit:} The lack of a universal protocol for TPC interoperability is hindering large-scale adoption, underscoring the need for standardization efforts. 

\section{Comparison between PLC and TPC}\label{Benchmark}

\begin{figure}[h]
\centering
\includegraphics[width=0.358\textwidth]{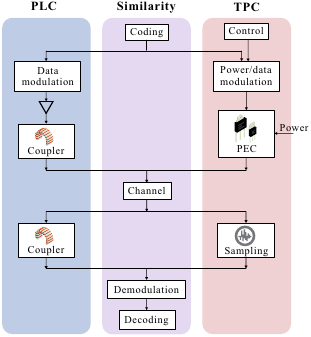}
\caption{Duality of PLC and TPC transmitter/receiver (Tx/Rx) structure in wired applications.}
\label{commonalities}
\end{figure}

As shown in Fig.~\ref{commonalities}, there are several structural similarities but also differences between PLC and TPC when considering only power line applications. 
This motivates a detailed comparison, which is performed next.  

\begin{table*}[h]
\caption{PLC vs.~TPC benchmarking. Strengths are marked by $+$ and bottlenecks by $-$.}
\centering
\begin{tabular}{l|l|l}
\hline
\textbf{Feature}             &\textbf{PLC}                                               & \textbf{TPC} \\
\hline
Hardware aspects             & {$-$ External modem on Tx and Rx sides}                   & {$+$ Applicable to any switched-mode PEC w/o extra Tx HW} \\
Data feed-in point           & {$+$ Data can be injected anywhere along power lines}     & {$-$ Data can be send by PECs only} \\
Data transmission method     & {$+$ Wide range of modulation schemes}                    & {$-$ Only 2-level switching signals possible} \\
Data rate                    & {$+$ $\sim$100~kbps to Gbps}                              & {$-$ $\sim$100~bps to Mbps} \\
Communication range          & {$+$ Very large range for NB-PLC}                         & {$-$ Short-range data transmission on busses} \\
Bidirectional communication  & {$+$ PLC-based duplex communication is common}            & {$-$ TPC-based duplex communication is under research} \\
Power quality                & {$-$ Significant ripple voltage}                          & {$+$ Trade-off between ripple voltage and data detection} \\
Voltage levels               & {$+$ Mostly applied for LV, rarely for MV and HV}         & {$-$ So far only tested for LV, but potential for MV and HV} \\
Main disturbances            & {$-$ Frequency-dependent propagation loss, impulse noise} & {$-$ Time-varying load randomly affects ripple voltage} \\
EMI concerns                 & {$-$ PLC emits radio waves}                               & {$+$ TPC operates at baseband} \\
Protocols and cyber security & {$+$ Standardized network protocols}                      & {$-$ Standardization is pending} \\
Latency                      & {$-$ Moderate delay}                                      & {$+$ Small delay possible} \\
Applications                 & {$-$ Only suitable for grid applications}                 & {$+$ Wired, wireless and optical applications} \\
\hline
\end{tabular}
\label{benchmarking}
\end{table*}

{\em Hardware aspects:} PLC requires an external transmitter, including in particular an amplifier and an inductive or capacitive coupling unit. TPC, on the other hand, does not require any additional PE components: the microcontroller intended for power control can usually also take on the task of switching pattern generation. This reduces complexity and hardware/installation/maintenance costs and increases reliability and sustainability, as the carbon footprint is lower. PLC and TPC receivers are similar, in both cases PE components are not required. Neither technique requires extra wiring. 

{\em Data feed-in point:} In the case of PLC, data can be injected anywhere along power lines, whereas in the case of TPC data can be send by PECs only. On the one hand, this offers PLC more flexibility. On the other hand, TPC is the more cost-effective solution.  

{\em Data transmission method:} PLC supports a wide range of symbol alphabets. In BB-PLC, up to 4096-QAM is used in conjunction with OFDM. High-frequency signals with large bandwidth are superimposed onto the power line. In contrast, with TPC only two-level switching signals are possible per half bridge. In addition, TPC is a baseband technology. The symbol duration is lower bounded by the switching period. Although the degrees of freedom for data transmission can be increased with more advanced PECs (such as multi-level, multi-modular and multi-port PECs), the cardinality of the symbol alphabet is likely to be lower than for PLC.

{\em Data rate:} Due to the large bandwidth in combination with the possibility of using higher-order modulation methods, the peak channel capacity of PLC is very large. The symbol rate is adjustable over a wide range. NB-PLC standards offer data rates of around 100~kbps and multiples thereof, while BB-PLC can achieve a very high throughput of up to 2~Gbps. In contrast, with TPC the symbol rate is coupled to the switching frequency and low-order modulation methods are common. Data rates from a few hundred bps up to about 1~Mbps have been reported. For both technologies, the channel capacity is time-varying due to the random nature of the channel distortions.

{\em Communication range:} The communication range strongly depends on signal bandwidth, error protection level, and channel conditions. As already elaborated, NB-PLC is in favor of large-distance communications. According to the current state-of-the-art, TPC is more useful for short-range communication, for instance between solar inverters and batteries. 

{\em Bidirectional communication:} Many PLC devices offer two-way data exchange. TPC-based duplex communication is currently under research.

{\em Power quality:} Any form of voltage superposition onto a power line negatively affects the power quality, energy efficiency, and network stability. There is a trade-off between power quality and ripple voltage. This holds for PLC as well as for TPC. PLC is superior in the sense that the bandwidth gap between control processes and communication signals is large, which simplifies filtering. However, TPC is superior in the sense that PEC topologies and modulation schemes exist where the ripple voltage is adjustable.

{\em Voltage levels:} While most PLC standards target LV lines, some support HV/MV grids. HV lines (110-380~kV) transmit PLC signals with low attenuation over long distances, while MV lines (10-30~kV) face higher attenuation due to more branches, but cover shorter spans. LV grids (110-400~V) pose the greatest path loss challenges, worsened by distribution transformers -- solved via bypass couplers, repeaters, or low-frequency PLC. NB-PLC on HV/MV lines is rare due to costly coupling and competition from fiber/wireless alternatives. So far, TPC has only been tested on LV grids.

{\em Main disturbances:} In PLC systems, frequency-selective propagation loss and impulse noise are key challenges, whereas in TPC systems, time-varying loads and connected PECs randomly affect the ripple voltage.

{\em EMI concerns:} PLC emits radio waves that interfere with other devices, while TPC operates at baseband. Random modulation and spread-spectrum are promising techniques to reduce radiation. Less regulatory constraints are expected for TPC. 

{\em Protocols and cyber security:} PLC is mature and secure, numerous international standards are available which also include encryption. In the case of TPC, however, network protocols and standardization are pending.

{\em Latency:} Low-delay communication is important in power control loops and machine-to-machine communication. In the case of PLC, the latency is dominated by the communication protocol. Vice versa, as protocols have not been established for TPC, latencies on the order of a few symbol durations are feasible for TPC systems with time-domain processing. 

{\em Applications:} Possible applications of PLC include but are not limited to \cite{Ndjiongue2019}  
\begin{itemize}
\item {\em Microgrid control:} PLC is able to provide distributed power control without additional wiring, thereby reducing latency.
\item {\em Smart grid applications:} PLC reduces operational costs by supporting advanced metering infrastructure, energy management, dynamic load balancing, distributed automation, fault detection, etc. 
\item {\em In-vehicle grids:} PLC enables simultaneous energy distribution and data transfer in cars, robots, busses, trains, ships, etc. Vehicular grids are usually DC grids. 
\item {\em Automation:} Home and industrial automation are important PLC applications for energy savings and a reduction of the carbon footprint, particularly if coupled with renewable energy sources like solar modules and energy storage. 
\item {\em Home access:} Internet access via the low-voltage distribution grid is an instance of last-mile access. As such, PLC is an alternative to DSL and WiMax. 
\item {\em Home-area networking:} In-house applications such as infotainment, multimedia distribution and surveillance are high-speed PLC applications. PLC may also provide a backbone for visible light communication, near field communication, and similar services. 
\item {\em Wide-area networking:} Potential PLC applications of wide-area networking are power control and telemetry, like transmission line inspection. 
\end{itemize}
Microgrid control, smart grid applications, vehicular grids, home and industrial automation, and home access are among the applications that PLC and TPC have in common. Additionally, there are several applications where TPC is a superior option in certain aspects or can not be implemented using conventional PLC:  
\begin{itemize}
\item {\em Self-diagnosis of power converters:} TPC makes it easy to broadcast converter-internal status information to the outside.
\item {\em Edge computing:} Powered converters with embedded communication are able to connect IoT devices in real time, also in energy-constraint environments.
\item {\em Cyber-physical power systems:} TPC is able to integrate physical power infrastructure with cyber components.
\item {\em Low-delay signaling:} At the PHY layer, TPC enables ultra-low-delay data transmission suitable for power control loops and machine-to-machine communication. 
\item {\em Battery management systems:} TPC enables real-time broadcasting of status parameters in battery modules.
\item {\em Motor control systems:} TPC has successfully been tested in switched reluctance machines and may be suitable for robotic assembly lines.
\item {\em Wireless applications:} Among potential wireless use cases of TPC are radio-frequency applications like sensor networks and optical applications like visible light communication. 
\end{itemize}

In TABLE~\ref{benchmarking}, the benchmarking metrics are summarized. In conclusion, PLC is better suited for long-range, high-data-rate applications where power and communication are separate concerns (e.\,g., grid monitoring or broadband over power lines). TPC is more efficient for short-range, low-complexity embedded communication within PE systems (e.\,g., localized grids and renewable energy systems), but also for wireless scenarios. The choice may depend on the targeted application. Future advancements in TPC, to be discussed next, may shorten the gap. 

\section{Future Research Directions}\label{Future}
While PLC is a mature technology \cite{Ndjiongue2019}, regarding TPC still many evolutionary steps need to be done. The following list addresses important aspects and may serve as a roadmap for future research. 

{\em Advanced PE converters:} The limited data rate can be boosted by advanced converter/inverter topologies and by fast wide-bandgap semiconductor switches. The increase in switching speeds is also advantageous in terms of weight and volume, but at the risk of electromagnetic radiation. 

{\em Channel characterization:} Widely recognized channel models are beneficial during research, testing, and standardization phases. Given a PEC topology, a channel model and a multiuser configuration, the channel capacity should be explored in order to obtain performance bounds and system design hints. 

{\em Artificial-intelligent-driven adaptive modulation and multiuser access (PHY to MAC layer):} Machine-learning-opti\-mized resource allocation and receiver architectures are promising to cope with random channel variations and load changes. 

{\em Routing:} Data and energy routing should be optimized simultaneously. This aspect deserves more research tailored to TPC. 

{\em Protocols and data security (upper layers):} Robust protocols that can be applied universally and sophisticated encryption are essential for critical scenarios.

{\em Networking:} The evolution from peer-to-peer scenarios to multiuser scenarios is important to create new use cases and a wider acceptance. 

{\em IoT and wireless sensor networks:} While initially intended for PLC-like applications, TPC should be adapted for use in IoT applications and sensor networks as well. 
As TPC supports localized intelligence, it facilitates the scalability of smart energy networks.  

{\em 6G convergence:} TPC promises ubiquitous connectivity in combination with power supply. It may support low-latency communication for grid-edge devices in 6G networks.

\begin{figure}[h]
\centering
\includegraphics[width=0.495\textwidth]{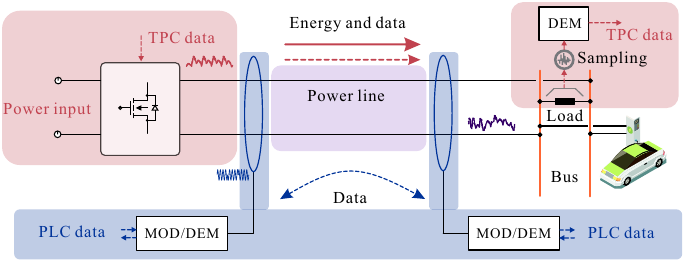}
\caption{Novel proposal of a hybrid system combining PLC and TPC. In this hybrid approach, data from PLC and TPC transmitters are send over the same power line. Either a common receiver or different receivers may be used.} 
\label{PLC-TPC}
\end{figure}

{\em Hybrid PLC-TPC networking:} Hybrid networks combining PLC and TPC (and possibly wireless devices) could enhance future energy-data networks, see Fig.~\ref{PLC-TPC}. Together with an abstraction layer such as IEEE 1905.1, a heterogeneous network could be created that combines TPC with PLC, Wi-Fi and Ethernet to foster the convergence of wired and wireless systems. 

{\em Standardization:} Standardization is the key to entering the mass market.

This roadmap illustrates that cross-disciplinary research on TPC is required at all levels, from layer 0 (converter topology and channel modeling) via the PHY and MAC layer (modulation, coding, multiple access, interference mitigation) and data/energy routing through protocol and cyber security aspects to the application layer, to convergence, and finally standardization. 

\section{Conclusion}\label{Conclusion}
PLC is a mature and widespread technology that uses existing electrical networks for digital communication. Meanwhile, TPC is a promising alternative by integrating data communication directly into the power conversion process. Each method has its own advantages and challenges regarding data rate, reliability, scalability, and cost. One of the biggest differences is that TPC can also be used in wireless and optical applications, making it the more universal concept. As smart grids, industrial automation, IoT and 6G evolve, the synergy between PLC and TPC could lead to more integrated, efficient, and resilient communication networks. Continued research is essential to optimize TPC designs and develop standards that can harness the full potential of this technology.

\bibliographystyle{IEEEtran}
\bibliography{IEEEabrv,references}

\end{document}